# Impurity Band Conduction in Si-doped $\beta$-Ga$_2$O$_3$ Films

Anil Kumar Rajapitamahuni[1, a], Laxman Raju Thoutam[1, 2], Praneeth Ranga[3], Sriram Krishnamoorthy[3], and Bharat Jalan[1, a]

[1]Department of Chemical Engineering and Materials Science, University of Minnesota, Minneapolis, MN, 55455

[2]Now at Department of Electronics and Communications Engineering, SR University, Warangal Urban, Telangana, India. 506371

[3]Department of Electrical and Computer Engineering, The University of Utah, Salt Lake City, UT, 84112

a) rajap016@umn.edu , bjalan@umn.edu

By combining temperature-dependent resistivity and Hall effect measurements, we investigate donor state energy in Si-doped $\beta$-Ga$_2$O$_3$ films grown using metal-organic vapor phase epitaxy (MOVPE). High magnetic field ($H$) Hall effect measurements (-90 kOe $\leq H \leq$ + 90 kOe) showed non-linear Hall resistance for T < 150 K revealing two-band conduction. Further analyses revealed carrier freeze-out characteristics in both bands yielding donor state energies of ~ 33.7 and ~ 45.6 meV. The former is consistent with the donor energy of Si in $\beta$-Ga$_2$O$_3$ whereas the latter suggests a residual donor state, likely associated with a DX center. This study provides a critical insight into the impurity band conduction and the defect energy states in $\beta$-Ga$_2$O$_3$ using high-field magnetotransport measurements.

$\beta$-Ga$_2$O$_3$ possesses wide bandgap (4.6 - 4.9 eV)[1], high theoretical electrical breakdown (~ 8 MV/cm)[2] and high conductivity with reasonably high room-temperature mobility[3], ~ 184 cm$^2$V$^{-1}$s$^{-1}$ making it an attractive candidate for high-power device applications.[4,5] Furthermore, access to low cost, large-scale (up to 6") native substrates with low threading dislocation density (10$^3$-10$^4$ cm$^{-2}$) offers significant advantages for $\beta$-Ga$_2$O$_3$ epitaxy.[5] $\beta$-Ga$_2$O$_3$ has monoclinic symmetry (space group C2/m, lattice parameters a = 12.214 Å, b = 3.0371 Å, c = 5.7981 Å, and $\beta$ = 103.83° and is the only stable polymorph of Ga$_2$O$_3$ up to the melting point.[6] Within the structure, Ga$^{3+}$ ions are both tetrahedrally and octahedrally coordinated, while O$^{2-}$ ions are either trigonally or tetrahedrally coordinated.[7] This structural complexity complicates the doping study. For instance, it is conceivable that the local electronic structure can vary significantly depending on the dopant size and the sites it occupies. Despite this obvious challenge, the thermal, optical, and electrical transport properties of $\beta$-Ga$_2$O$_3$ have been studied extensively, both experimentally and using first-principles calculations[2-6,8-21]

Silicon (Si) is shown to be a shallow n-type donor in $\beta$-Ga$_2$O$_3$.[5,6] Yet, there remains a large inconsistency in the reported ionization energy of Si. For instance, activation energy of Si in Si-doped $\beta$-Ga$_2$O$_3$ ranges from 16 to 50 meV.[10,22] The variation in donor activation energies has been attributed to the donor density[22] and to the presence of defects and impurities arising from various growth techniques.[23,24] Relatively deeper donors with activation energies 80-120 meV have also been reported, the origin of which are attributed to the presence of antisites, interstitials and/or extrinsic impurities.[3,17,25] Deep level states such as DX centers, which are defect complexes formed between isolated substitutional donor atom (D) and an unknown lattice defect (X) are also studied in $\beta$-Ga$_2$O$_3$. Using electron paramagnetic resonance (EPR) study, Song et al reported DX center in unintentionally doped $\beta$-Ga$_2$O$_3$ with activation energies, 44 - 49 meV for partially activated centers, reducing to 17 meV for a fully activated DX centers.[26] However, recent transport measurements refuted the presence of DX centers in doped $\beta$-Ga$_2$O$_3$ based on the low-field magnetotransport analysis.[22] As presence of DX centers is determinantal for Ga$_2$O$_3$ based heterojunction devices, this certainly raises important questions: why there is such discrepancy discrepancy in the reports of DX centers in $\beta$-Ga$_2$O$_3$? Can this be due to the variation in the materials depending on the synthesis conditions? Clearly, further investigations of the growth condition-structure-defect-property relationships would help address these questions.

In an attempt to investigate donor state energies in Si-doped $\beta$-Ga$_2$O$_3$, we performed a detailed temperature-dependent magnetotransport studies of homoepitaxial Si-doped $\beta$-Ga$_2$O$_3$ films grown via metal-organic vapor phase epitaxy (MOVPE). Low magnetic field ($H$) Hall effect measurements (-20 kOe $\leq H \leq$ + 20 kOe) showed single band conduction with an activation energy of ~ 17 meV. In sharp contrast, high-magnetic field (-90 kOe $\leq H \leq$ + 90 kOe) Hall effect measurements revealed two-band conduction with activation energies, ~ 34 and ~ 46 meV. We discuss the origin of these energy states in the context of Si donor state energy and a residual donor state, possibly a DX$^-$ center respectively.

Si-doped $\beta$-Ga$_2$O$_3$ films were grown on (010) Fe-doped semi-insulating $\beta$-Ga$_2$O$_3$ substrates using MOVPE reactor (Agnitron Agilis). Triethylgallium (TEGa), molecular O$_2$ were used as a source of Ga and oxygen in the presence of Ar as a carrier gas. Substrate temperature was fixed at 810 °C. Si was used as n-type dopant and was controlled by varying the molar ratio of diluted silane (SiH$_4$) to TEGa ratio.[13] Ohmic contacts were achieved by sputtering Ti/Au (50 nm/50 nm) stacks using shadow mask followed by a rapid thermal annealing at 470 °C in nitrogen for 90 secs. Temperature-dependent electrical measurements were performed in Van der Pauw geometry using a physical property measurement system (PPMS® DynaCool™). Excitation currents of 1 – 10 μA were used.

Figure 1a and 1b show temperature-dependent resistivity ($\rho$) and carrier density respectively from a 655 nm Si-doped $\beta$-Ga$_2$O$_3$/Fe-doped $\beta$-Ga$_2$O$_3$ (010). Schematic of the sample structure is shown in the inset. It is noted that Hall measurement in the low-field between ± 20 kOe yielded linear behavior. Hall coefficient ($R_H$) in figure 1b is therefore extracted from the linear slope of Hall resistance ($R_{xy}$) vs. $H$ where $H$ was varied between ± 20 kOe. Temperature dependence carrier density showed a decrease in carrier density from $7.8 \times 10^{17}$ cm$^{-3}$ at 300 K to $6.74 \times 10^{16}$ cm$^{-3}$ at 65 K followed by an unexpected upturn at low temperatures, $40 \leq T \leq 65$ K. To further elucidate this observation, we show in figures 1c and 1d an Arrhenius plots of $\rho$ and $R_H$ revealing nominally three distinct regimes: (i) $225 \leq T \leq 300$ K where $\rho$ decreases and $R_H$ increases with decreasing temperature; (ii) $65$ K $\leq T \leq 225$ K, where there is a rapid increase in $\rho$ with decreasing temperature accompanied by an increase in $R_H$; and (iii) $40 \leq T \leq 65$ K where $\rho$ continues to increase with decreasing temperature but now $R_H$ begins to decrease. This behavior is remarkably similar to the previously observed temperature-dependence of $\rho$ and $R_H$ in doped Germanium (Ge) and other heavily-doped semiconductors.[27,28] These characteristics have further been attributed to impurity band conduction where electrons move in both conduction and an impurity band. Most recently, Kabilova et.al. also observed an identical behavior in Sn-doped $\beta$-Ga$_2$O$_3$ and attributed it to the two-band conduction.[29] At higher T, conduction is dominated by electrons in the conduction band whereas at low temperatures donor-derived impurity band conduction dominates.[28] Given two-band conduction, one can therefore write the overall resistivity and Hall coefficient ($R_H$) as

$$\rho(T) = t_{film}(n_1 e \mu_1 + n_2 e \mu_2)^{-1} \qquad (1)$$

$$R_H = (n_1\mu_1^2 + n_2\mu_2^2)\,(e\,(n_1\mu_1 + n_2\mu_2)^2)^{-1} \qquad (2)$$

where ($n_1$, $\mu_1$) and ($n_2$, $\mu_2$) represent temperature-dependent sheet carrier density and mobility in conduction band and impurity band respectively. $t_{film}$ represents the film thickness.

To further investigate two-band conduction in our films, we performed high-field Hall measurements. Figure 2a shows $R_{xy}$ as a function of $H$ at 40 K ≤ T < 150 K. Metal contacts became non-ohmic at T < 40 K preventing lower temperature measurements. $H$ was swept between ± 90 kOe. Longitudinal resistance ($R_{xx}$) as a function of $H$ is shown in Figure S1 revealing a positive magnetoresistance behavior at all temperatures whereas $R_{xy}$ ($H$) showed non-linearity as illustrated in figure 2a. The latter is consistent with two-band conduction. We analyzed our experimental results using two-band conduction model. In this model, $R_{xy}$ ($H$) can be written as:

$$R_{xy}(H) = \frac{-(H/e)[(n_1\mu_1^2 + n_2\mu_2^2) + H^2\mu_1^2\mu_2^2(n_1+n_2)]}{[(n_1\mu_1 + n_2\mu_2)^2 + H^2\mu_1^2\mu_2^2(n_1+n_2)^2]} \qquad (3)$$

In this equation, there are four unknowns ($n_1$, $\mu_1$ $n_2$, and $\mu_2$) that can be further reduced to two unknowns by calculating Hall conductance, $G_{xy}(H)$ using experimentally measured $R_{xy}(H)$ and $R_{xx}(H)$.

$$G_{xy}(H) = -\frac{R_{xy}}{(R_{xy}^2 + R_{xx}^2)} = eH\left(\frac{(C_1\mu_1 - C_2)}{(\mu_1/\mu_2 - 1)(1+\mu_2^2 B^2)} + \frac{(C_1\mu_2 - C_2)}{(\mu_2/\mu_1 - 1)(1+\mu_1^2 B^2)}\right) \qquad (4)$$

Where $C_1 = n_1\mu_1 + n_2\mu_2$ and $C_2 = n_1\mu_1^2 + n_2\mu_2^2$. It should be noted that $C_1$ and $C_2$ are known experimentally from the conductance and the linear slope of the Hall conductance respectively at zero magnetic field. Details of this analysis can be found elsewhere.[30] Figure 2b shows calculated $G_{xy}(H)$ along with fits (solid lines) using equation (4) at different temperatures revealing excellent match between experiments and two-band conduction model. This analysis yielded µ₁ and µ₂ (from the fits), which in turns allowed us to calculate n₁ and n₂.[30]

Figure 3 shows T-dependent $n_1$, $\mu_1$, $n_2$, and $\mu_2$ at 40 K ≤ T ≤ 150 K. Using $n_1$, $\mu_1$ $n_2$, and $\mu_2$ as a function of T, we calculated $\rho$ and $R_H$ using equation (1) and (2). The calculated $\rho$ (T) and $R_H$ (T) is shown in figure 1c and 1d using open red symbols revealing excellent match with experimental data. Our analysis therefore shows self-consistent results providing further confidence in the two-band conduction model. It is further noted that our analysis yielded reasonably good fits with similar values of $\mu_1$ and for a range of $\mu_2$ values between 0.1 and 10 cm$^2$V$^{-1}$s$^{-1}$ (shown as a shaded region in figure 3a). Noticeably, $\mu_2$ has a significantly lower value as one would expect from an impurity band conduction. In figure 3a, we only show $\mu_2$ = 10 cm$^2$V$^{-1}$s$^{-1}$, which is closer to the mobility values reported for impurity band conduction in $\beta$-Ga$_2$O$_3$.[10,12,31] On the other hand, $\mu_1$ first increases with decreasing temperature, reaching a peak value of 796 cm$^2$V$^{-1}$s$^{-1}$ at 65 K, and then begins to decrease. The increase in $\mu_1$ follows $T^{-0.5}$ behavior (figure 3c) which is consistent with phonon-related scattering in $\beta$-Ga$_2$O$_3$ in agreement with the previous reports.[11,22] Whereas the drop in mobility for T < 65 K is consistent with the ionized impurity scattering. Significantly, this temperature is same at which $R_H$ was found to decrease in figure 1d suggesting the scattering centers are likely the donor-derived ionized impurities. Unlike $\mu_1$, $\mu_2$ was found to be low and T-independent which is again consistent with the presence of impurity band.[10,12]

We now turn to the discussion of donor state energy. First, we present results from the analyses of low-field Hall effect measurements yielding single band conduction with an activation energy of ~ 17 meV (Figure S2). This energy state is in good agreement with the published results of Si activation energy in Si-doped $\beta$-Ga$_2$O$_3$ near the Mott insulator-to metal transition.[22] However, our analyses using high-field Hall effect measurements resulted in a deeper understanding of transport activation behavior. Figure 3b shows Arrhenius plots for $n_1^{3D}$ ($n_1/t_{film}$) and $n_2^{3D}$ ($n_2/t_{film}$) extracted from the two-band conduction model. This plot yielded linear slopes with activation energies of $E_a^{n_1}$ = 33.7 meV and $E_a^{n_2}$ = 11.9 meV respectively. The donor ionization energy of Si in *β*-Ga$_2$O$_3$ is reported to be ~ 36 meV, which is close to $E_a^{n_1}$ suggesting Si shallow donors are the source of high-mobility carriers and that they are responsible for conduction at higher temperatures.[10,22] The corresponding high mobilities, 285 cm$^2$V$^{-1}$s$^{-1}$ (T = 150 K) < $\mu_1$ < 678 cm$^2$V$^{-1}$s$^{-1}$ (40 K) further corroborates with the transport occurring in the conduction band. In addition, we found a residual donor state ~ 12 meV lying below the primary Si donor state, as shown schematically in the inset of figure 3b, with a donor state energy of 45.6 meV (= 33.7 + 11.9 meV). The donor state energy of 45.6 meV is consistent with the DX center in *β*-Ga$_2$O$_3$, which is estimated to be 44 – 49 meV using EPR measurements.[26] To the best of our knowledge, this is the first evidence of the possible DX center in *β*-Ga$_2$O$_3$ using high-field magnetotransport measurements. We however note that while this study provides a clear evidence of a residual donor state at ~ 46 meV, which may likely be associated with a DX center, it is non-trivial to assign it to a specific defect-type. Whether this is DX center, or a new defect-complex requires a detailed study of film growth combined with high-field transport and spectroscopy characterizations. Future study should be directed to investigate the relationship between synthesis conditions and defect formation in *β*-Ga$_2$O$_3$.

In summary, we have investigated donor state energy in doped *β*-Ga$_2$O$_3$ films via temperature-dependent resistivity and Hall effect measurements. Two-band conduction model described experimental data in addition to yielding donor state energies, ~ 34 meV and ~ 46 meV, which we attribute to Si donor and a potential DX center, respectively. In contrast, low-field transport yielded only one carrier type with an activation energy of ~ 17 meV in agreement with the published results. Our work provides new insights into the nature of the donor types in Si-doped $\beta$-Ga$_2$O$_3$ with implications in the development of high-power electronic devices.

**Acknowledgements:**

This work was supported primarily by the National Science Foundation through the University of Minnesota MRSEC under Award Number DMR-2011401. Part of this work was supported through the Air Force Office of Scientific Research (AFOSR) through Grant FA9550-19-1-0245 and through DMR-1741801. Portions of this work were conducted in the Minnesota Nano Center, which is supported by the National Science Foundation (NSF) through the National Nano Coordinated Infrastructure Network (NNCI) under Award Number ECCS-1542202. Part of this work was also carried out in the College of Science and Engineering Characterization Facility, University of Minnesota, which has received capital equipment funding from the NSF through the UMN MRSEC program. Thin film synthesis work at the University of Utah is supported primarily by the Air Force Office of Scientific Research under award number FA9550-18-1-0507 monitored by Dr Ali Sayir. Any opinions, findings and conclusions or recommendations expressed in this material are those of the authors and do not necessarily reflect the views of the United States Air Force. Material synthesis effort at the U of Utah also acknowledge support from National Science Foundation (NSF) under Award Number DMR-1931652. Part of the work was performed at the Utah Nanofab sponsored by the College of Engineering and the Office of the Vice President for Research.

**Data Availability**

The data that support the findings of this study are available from the corresponding author upon reasonable request.

**References**

1       H. H. Tippins,  Physical Review **140** (1A), A316 (1965).
2       Masataka Higashiwaki, Kohei Sasaki, Akito Kuramata, Takekazu Masui, and Shigenobu Yamakoshi,  Applied Physics Letters **100** (1) (2012).
3       Zixuan Feng, A. F. M. Anhar Uddin Bhuiyan, Md Rezaul Karim, and Hongping Zhao,  Appl. Phys. Lett. **114** (25) (2019).
4       Masataka Higashiwaki, Kohei Sasaki, Hisashi Murakami, Yoshinao Kumagai, Akinori Koukitu, Akito Kuramata, Takekazu Masui, and Shigenobu Yamakoshi,  Semiconductor Science and Technology **31** (3) (2016).
5       Jiaye Zhang, Jueli Shi, Dong-Chen Qi, Lang Chen, and Kelvin H. L. Zhang,  APL Materials **8** (2) (2020).
6       S.I. Stepanov, V. I. Nikolaev, V. E. Bougrov, and A. E. Romanov,  Rev. Adv. Mater. Sci. **44**, 63 (2016).
7       S. Geller,  J. Chem. Phys. **33**, 676 (1960).
8       Andrew J. Green, Kelson D. Chabak, Eric R. Heller, Robert C. Fitch, Michele Baldini, Andreas Fiedler, Klaus Irmscher, Gunter Wagner, Zbigniew Galazka, Stephen E. Tetlak, Antonio Crespo, Kevin Leedy, and Gregg H. Jessen,  IEEE Electron Device Letters **37** (7), 902 (2016).
9       Masataka Higashiwaki, Kohei Sasaki, Takafumi Kamimura, Man Hoi Wong, Daivasigamani Krishnamurthy, Akito Kuramata, Takekazu Masui, and Shigenobu Yamakoshi,  Applied Physics Letters **103** (12) (2013).
10      K. Irmscher, Z. Galazka, M. Pietsch, R. Uecker, and R. Fornari,  Journal of Applied Physics **110** (6) (2011).
11      Nan Ma, Nicholas Tanen, Amit Verma, Zhi Guo, Tengfei Luo, Huili Xing, and Debdeep Jena,  Applied Physics Letters **109** (21) (2016).
12      Toshiyuki Oishi, Yuta Koga, Kazuya Harada, and Makoto Kasu,  Applied Physics Express **8** (3) (2015).
13      Praneeth Ranga, Ashwin Rishinaramangalam, Joel Varley, Arkka Bhattacharyya, Daniel Feezell, and Sriram Krishnamoorthy,  Applied Physics Express **12** (11) (2019).
14      Kohei Sasaki, Masataka Higashiwaki, Akito Kuramata, Takekazu Masui, and Shigenobu Yamakoshi,  Journal of Crystal Growth **392**, 30 (2014).
15      J. B. Varley, J. R. Weber, A. Janotti, and C. G. Van de Walle,  Applied Physics Letters **97** (14) (2010).
16      Encarnación G. Víllora, Kiyoshi Shimamura, Takekazu Ujiie, and Kazuo Aoki,  Applied Physics Letters **92** (20) (2008).
17      Yuewei Zhang, Fikadu Alema, Akhil Mauze, Onur S. Koksaldi, Ross Miller, Andrei Osinsky, and James S. Speck,  APL Materials **7** (2) (2019).
18      Hong Zhou, Kerry Maize, Gang Qiu, Ali Shakouri, and Peide D. Ye,  Applied Physics Letters **111** (9) (2017).
19      Fabio Orlandi, Francesco Mezzadri, Gianluca Calestani, Francesco Boschi, and Roberto Fornari,  Applied Physics Express **8** (11) (2015).
20      K. Ghosh and U. Singisetti,  Int. J. High Speed Electron. Syst. **28**, 1940008 (2019).
21      Zixuan Feng, A F M Anhar Uddin Bhuiyan, Zhanbo Xia, Wyatt Moore, Zhaoying Chen, Joe F. McGlone, David R. Daughton, Aaron R. Arehart, Steven A. Ringel, Siddharth Rajan, and Hongping Zhao,  Phys. Status Solidi RRL **14**, 2000145 (2020).

22   Adam T. Neal, Shin Mou, Subrina Rafique, Hongping Zhao, Elaheh Ahmadi, James S. Speck, Kevin T. Stevens, John D. Blevins, Darren B. Thomson, Neil Moser, Kelson D. Chabak, and Gregg H. Jessen,  Applied Physics Letters **113** (6) (2018).
23   Stephan Lany,  APL Materials **6** (4) (2018).
24   Matthew D. McCluskey,  Journal of Applied Physics **127** (10) (2020).
25   A. T. Neal, S. Mou, R. Lopez, J. V. Li, D. B. Thomson, K. D. Chabak, and G. H. Jessen, Sci Rep **7** (1), 13218 (2017).
26   N. T. Son, K. Goto, K. Nomura, Q. T. Thieu, R. Togashi, H. Murakami, Y. Kumagai, A. Kuramata, M. Higashiwaki, A. Koukitu, S. Yamakoshi, B. Monemar, and E. Janzén,  J. Appl. Phys. **120** (23) (2016).
27   H. Fritzsche and M. Cuevas,  Phys. Rev. **119** (4), 1238 (1960).
28   N. F. Mott and W. D. Twose,  Advances in Physics **10** (38), 107 (1961).
29   Z. Kabilova, C. Kurdak, and R. L Peterson,  Semicond. Sci. Technol. **34**, 03LT02 (2019).
30   N. Bansal, Y. S. Kim, M. Brahlek, E. Edrey, and S. Oh,  Phys Rev Lett **109** (11), 116804 (2012).
31   Toshiyuki Oishi, Kazuya Harada, Yuta Koga, and Makoto Kasu,  Japanese Journal of Applied Physics **55** (3) (2016).

**Figures (Color Online):**

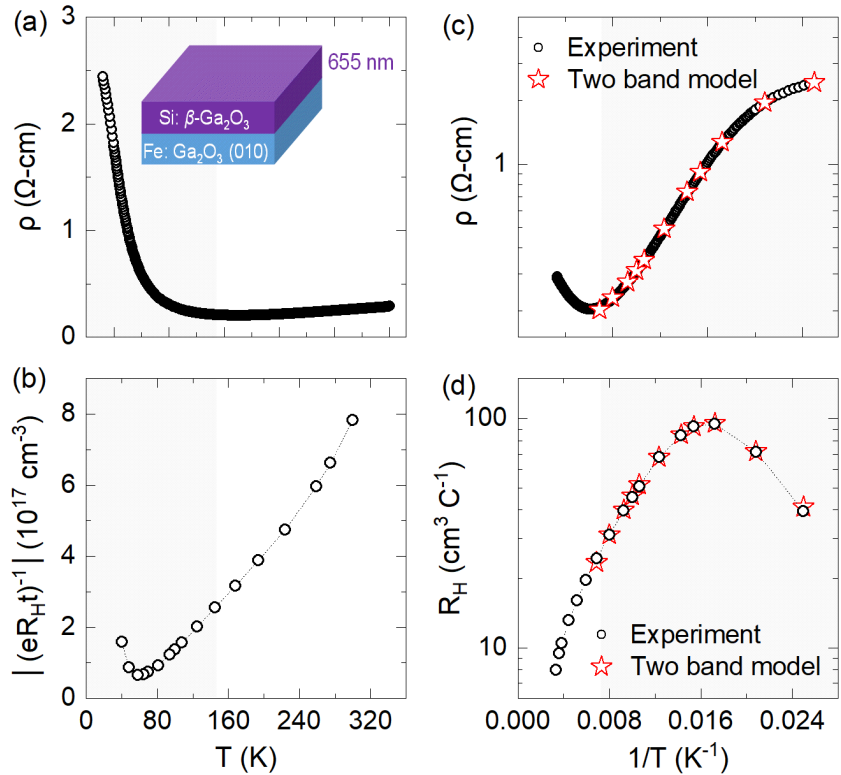

**Figure 1:** (a) Temperature-dependent $\rho$ and $R_H$ from a 655 nm Si-doped $\beta$-Ga$_2$O$_3$/Fe-doped $\beta$-Ga$_2$O$_3$ (010). Inset shows a schematic of sample structure. (b) 3D carrier density $(eR_Ht)^{-1}$ as a function of temperature, where $R_H$ is Hall coefficient, t is film thickness and e is an electronic charge. (c-d) Arrhenius plots of $\rho$ and $R_H$. The red symbol in part (c) and (d) are calculated $\rho$ and $R_H$ using two-band conduction model. Grey shaded region highlights the temperature range in which activated transport occurs.

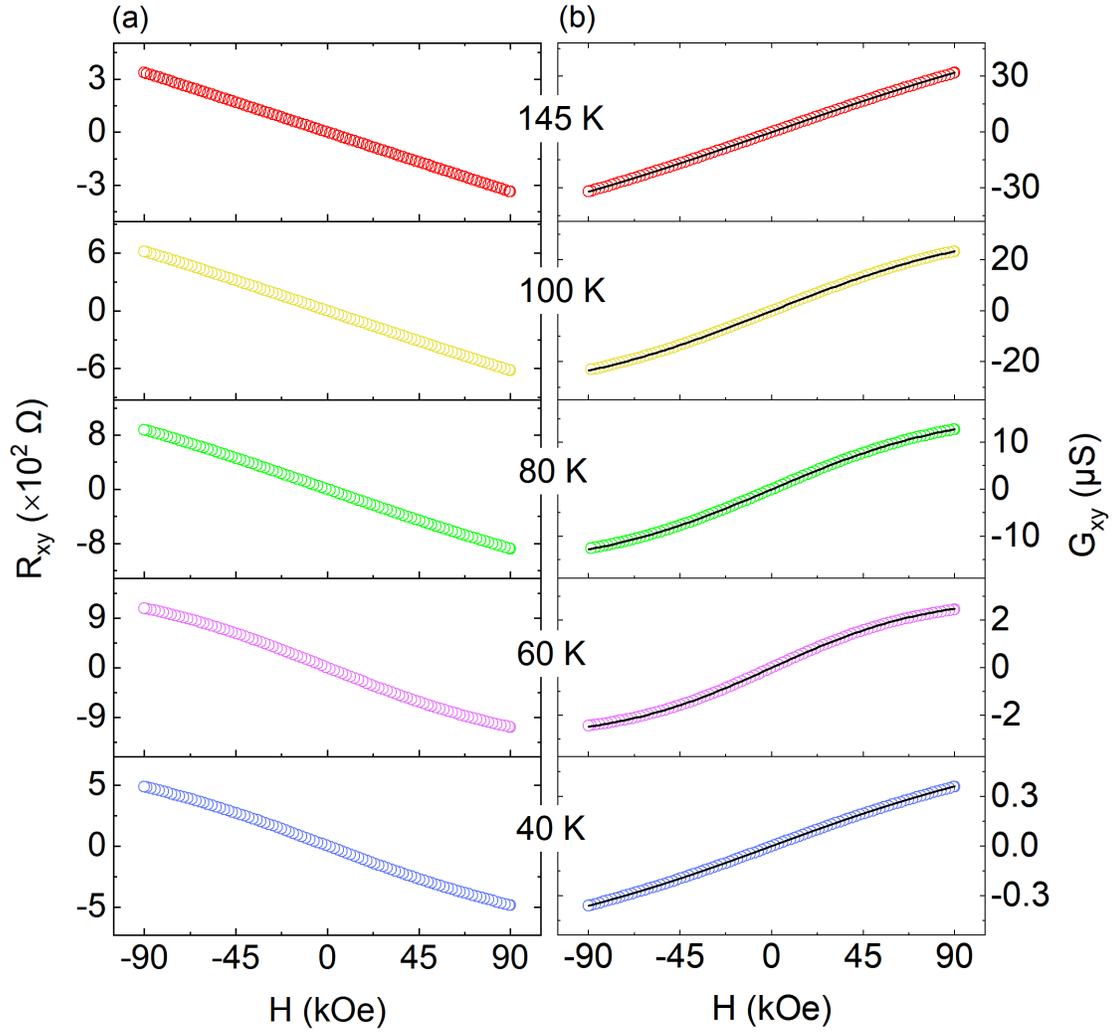

**Figure 2:** (a) $R_{xy}$ vs $H$ at T < 150 K showing non-linear behavior. (b) Hall conductance, $G_{xy} = -R_{xy}/(R_{xy}^2 + R_{xx}^2)$, along with fits (black solid lines) using two-band conduction model.

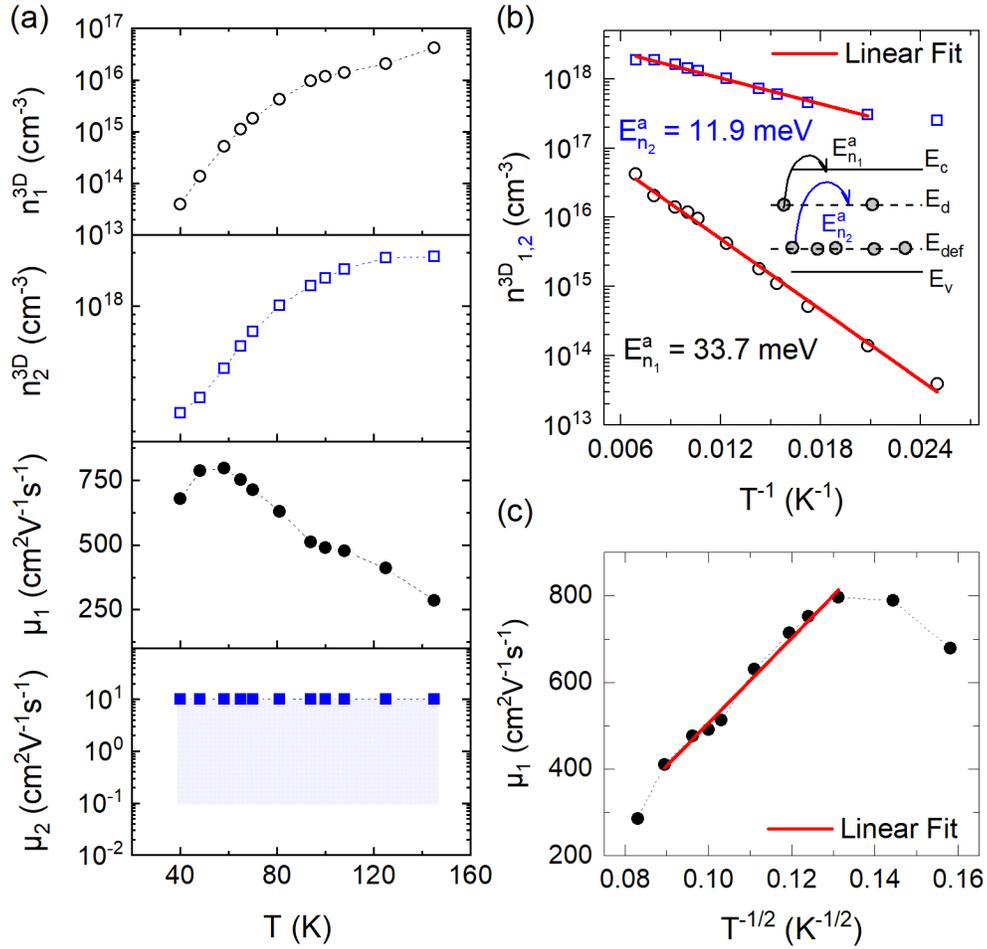

**Figure 3:** (a) 3D carrier densities, $n_1^{3D}$ and $n_2^{3D}$ and their corresponding mobilities $\mu_1$ and $\mu_2$ extracted from the two-band conduction model as a function of temperature. A shaded region in $\mu_2$ vs T plot shows a range of possible $\mu_2$ values. (b) Arrhenius plots with linear fits for $n_1^{3D}$ and $n_2^{3D}$ with corresponding activation energies $E_{n1}^a$ and $E_{n2}^a$. Inset shows defect state energies illustrating $E_{n1}^a$ and $E_{n2}^a$. Here, $E_d$ and $E_{def}$ refer to Si donor state and residual state energies respectively within the bandgap. (c) $\mu_1$ vs $T^{-1/2}$ along with a linear fit. The dashed lines are guide to the eye.

# Supplementary Information

## Impurity Band Conduction in Si-doped $\beta$-Ga$_2$O$_3$ Films

Anil Kumar Rajapitamahuni[1, a], Laxman Raju Thoutam[1, 2], Praneeth Ranga[3], Sriram Krishnamoorthy[3], and Bharat Jalan[1, a]

[1]Department of Chemical Engineering and Materials Science, University of Minnesota, Minneapolis, MN, 55455

[2]Now at Department of Electronics and Communications Engineering, SR University, Warangal Urban, Telangana, India. 506371

[3]Department of Electrical and Computer Engineering, The University of Utah, Salt Lake City, UT, 84112

a) rajap016@umn.edu , bjalan@umn.edu

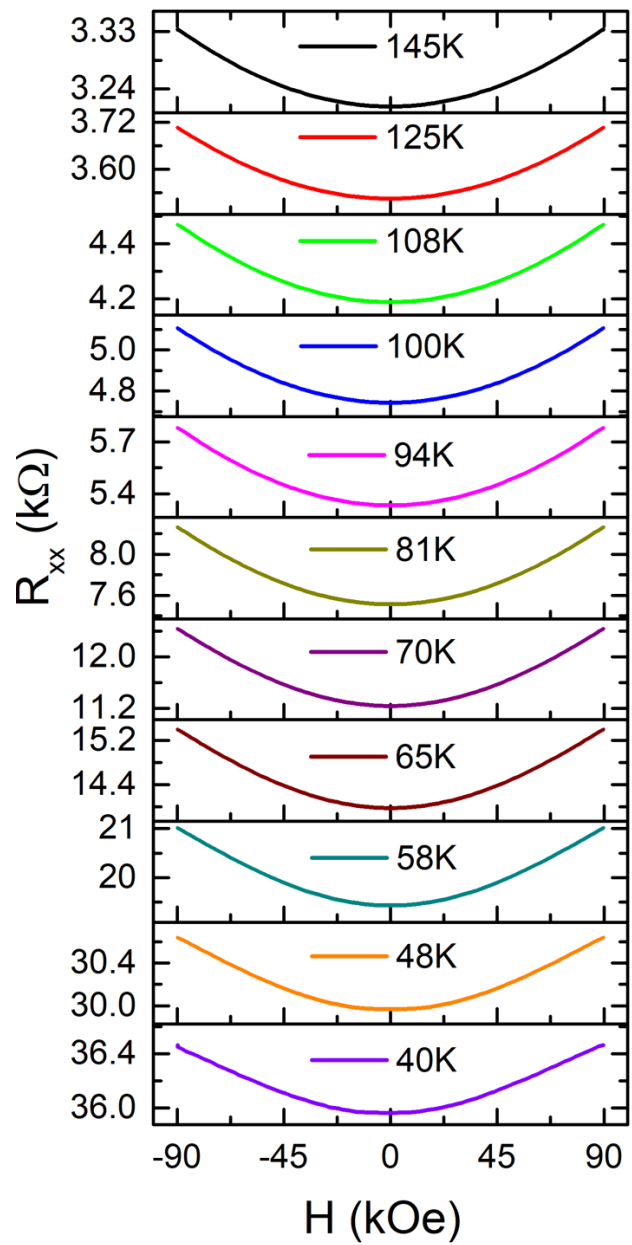

**Figure S1:** Longitudinal resistance ($R_{xx}$) as a function of *H* (-90 kOe T ≤ H ≤ +90 kOe) at 40 K ≤ T < 150 K showing positive magnetoresistance.

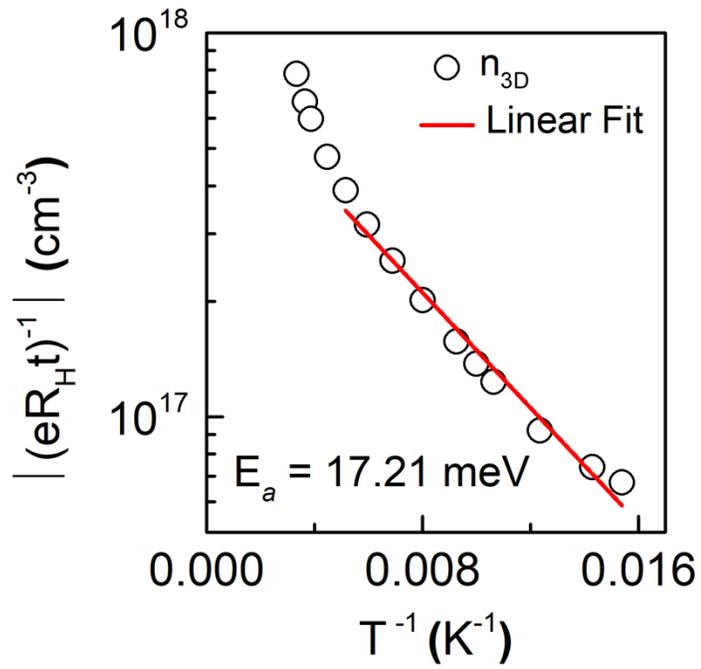

**Figure S2:** Arrhenius plot of carrier density as a function of temperature. Carrier density was calculated using low-field (- 20 kOe T ≤ H ≤ + 20 kOe) Hall effect measurements which yielded linear Hall slope.